\documentclass[pdflatex,sn-mathphys-num]{sn-jnl}

\usepackage{graphicx}%
\usepackage{multirow}%
\usepackage{amsmath,amssymb,amsfonts}%
\usepackage{amsthm}%
\usepackage{mathrsfs}%
\usepackage[title]{appendix}%
\usepackage{xcolor}%
\usepackage{textcomp}%
\usepackage{manyfoot}%
\usepackage{booktabs}%
\usepackage{algorithm}%
\usepackage{algorithmicx}%
\usepackage{algpseudocode}%
\usepackage{listings}%
\usepackage{siunitx}

\theoremstyle{thmstyleone}%
\theoremstyle{thmstyletwo}%

\theoremstyle{thmstylethree}%

\usepackage{titlesec}
\titleformat{\section}{\normalfont\large\bfseries}{ }{0pt}{}
\titleformat{\subsection}{\normalfont\normalsize\bfseries}{ }{0pt}{}
\begin{document}

\title[Article Title]{2D abrupt nano-junctions blending sp-sp$^2$ bonds on atomically precise heterostructures
} 

\author[1]{\fnm{Alice} \sur{Cartoceti}}\email{alice.cartoceti@polimi.it}
\equalcont{These authors contributed equally to this work.}

\author[2,3]{\fnm{Simona} \sur{Achilli}}\email{simona.achilli@unimi.it}
\equalcont{These authors contributed equally to this work.}

\author[2]{\fnm{Masoumeh} \sur{Alihosseini}}

\author[4]{\fnm{Adriana E.} \sur{Candia}}

\author[2]{\fnm{Enrico} \sur{Beltrami}}

\author[1]{\fnm{Paolo} \sur{D'Agosta}}

\author[5]{\fnm{Alessio} \sur{Orbelli Biroli}}

\author[6]{\fnm{Francesco} \sur{Sedona}}

\author[1]{\fnm{Andrea} \sur{Li Bassi}}

\author[7,8]{\fnm{Jorge} \sur{Lobo Checa}}

\author[1]{\fnm{Carlo S.} \sur{Casari}}

\affil*[1]{\orgdiv{Department of Energy}, \orgname{Politecnico di Milano}, \orgaddress{\street{via Lambruschini 6}, \city{Milano}, \postcode{20156}, \state{Italy}}}

\affil[2]{\orgdiv{Department of Physics ‘Aldo Pontremoli'}, \orgname{Università degli Studi di Milano}, \orgaddress{\street{Via G. Celoria 16}, \city{Milano}, \postcode{20133}, \state{Italy}}}

\affil[3]{\orgdiv{INFN Sezione di Milano and ‘European Theoretical Spectroscopy Facility' (ETSF)}, \orgaddress{\street{Via G. Celoria 16}, \city{Milano}, \postcode{20133}, \state{Italy}}}

\affil[4]{\orgdiv{Instituto de Física del Litoral}, \orgname{ Consejo Nacional de Investigaciones Científicas y Técnicas, Universidad Nacional del Litoral (IFIS-Litoral, CONICET-UNL)}, \orgaddress{\city{Santa Fe}, \postcode{3000}, \state{Argentina}}}

\affil[5]{\orgdiv{Department of Chemistry}, \orgname{Università di Pavia}, \orgaddress{\street{Via Taramelli 12}, \city{Pavia}, \postcode{27100}, \state{Italy}}}

\affil[6]{\orgdiv{Dipartimento di Scienze Chimiche}, \orgname{Università Degli Studi Di Padova}, \orgaddress{\city{Padova}, \postcode{35131}, \state{Italy}}}

\affil[7]{\orgdiv{Instituto de Nanociencia y Materiales de Aragon (INMA)}, \orgname{CSIC-Universidad de Zaragoza}, \orgaddress{\city{Zaragoza}, \postcode{50009}, \state{Spain}}}

\affil[8]{\orgdiv{Departamento de Física de la Materia Condensada}, \orgname{Universidad de Zaragoza}, \orgaddress{\city{Zaragoza}, \postcode{50009}, \state{Spain}}}

\abstract  {Two-dimensional heterostructures combining sp-sp$^2$ hybridization—blending graphene with graphyne-based allotropes—offer substantial potential for enhancing the tunability of electronic and transport properties while providing significant structural flexibility. These attributes are desirable for next generation nanoscale electronic applications. Despite such potential, their experimental realization remains elusive, as synthesized carbon heterostructures are limited to doped, graphene-based systems exhibiting exclusively sp$^2$ hybridization. Here, we demonstrate the on-surface synthesis of covalently bonded sp-sp$^2$ lateral heterostructures between graphene nanoribbons and graphdiyne networks on Au(111). Atomic-resolution scanning tunnelling microscopy, combined with density functional theory, reveals the formation mechanism of the covalent interfacial bonds between nanoribbons and graphdiynes, also highlighting the key role of surface chemistry. Bromine atoms deriving from the molecules dehalogenation and chemisorbed along the nanoribbon inhibit the junction formation, but bonding efficiency can be boosted up to 71\% by controlled removal of these by-products. Electronic structure and transport calculations show that the 2D heterostructure by itself is characterized by disentangled properties for the two subsystems, forming an atomically narrow junction enabling voltage-tunable spatial current separation in two dimensions. There results define a viable strategy for engineering graphene-based sp-sp$^2$ heterostructures, paving the way for the design and synthesis of all-carbon nanoscale electronic architectures.}


\maketitle

\section{Introduction}\label{sec1}

Heterostructures can be considered as the building blocks for modern semiconductor physics \cite{alferov2001nobel}, being central to the development of high-performance electronic and optoelectronic devices. The continued demand for dimensional downscaling and improved performance \cite{radsar2021graphene} has led to the development of two-dimensional (2D) heterostructures, constituted by vertically or laterally stitched atomically thin materials interacting through Van der Waals forces or covalent bonds. Lateral 2D heterostructures are typically obtained through bottom-up approaches, and recent publications have shown the potential of atomically abrupt interfaces in organic-inorganic lateral heterojunctions for next-generation atomic-scale circuitry \cite{liu2017self,zhang2017carbon}. 

Among 2D materials, graphene stands out for its huge electron mobility and transport properties \cite{castro2009electronic,schwierz2010graphene}. However, its semimetallic character and the absence of an intrinsic band gap severely limit its direct applicability in semiconducting devices. While band-gap engineering in graphene remains challenging, graphene-based heterostructures incorporating complementary 2D semiconductors open the way to the development of efficient devices \cite{liao2019van}.

In this context, the realization of all-carbon heterostructures integrating 2D carbon materials with tunable electronic properties complementing those of graphene would be highly desirable, also in the perspective of all-carbon electronics.

Graphdiynes (GDYs) have recently emerged as a promising class of atomically thin carbon materials with mixed $sp$-$sp^2$ hybridization, showing distinctive electronic and transport properties \cite{baughman1987structure,rabia2020structural, xue20182d, huang2015graphdiyne, li2019surface, gao2019graphdiyne, li2014graphdiyne,
casari2016carbon, serafini2021designing, zhao2024transmetalation, yang2020metalated}. The presence of diacetylenic bonds connecting benzene rings allows to overcome the zero-gap limitation of graphene, resulting in a new class of 2D carbon materials \cite{jia2017synthesis,yu2019graphdiyne}.

Atomically precise GDY networks \cite{sun2016dehalogenative,yang2020metalated}, as well as  GNRs \cite{lou2024graphene,radsar2021graphene,mateo2020surface,merino2017width,
ma2012first,li2018modular,chen2013tuning,wagner2013band,wakabayashi2009electronic,scuseria2006electronic,nakada1996edge}, can be obtained via on-surface synthesis (OSS).
This technique exploits the catalytic role of noble metal surfaces to induce dehalogenation and covalent coupling of specifically selected halogen-terminated molecular precursors, producing low-dimensional carbon nanostructures, thus representing the ideal approach for the bottom-up synthesis of hybrid carbon-based 2D interfaces \cite{sun2016dehalogenative,hu2017recent,bjork2013mechanisms,bjork2016reaction,zhou2017surface,li2024theoretical,wang2024universal,cai2010atomically, cartoceti2025surface}.

Herein, we report on the OSS of a covalently bonded lateral heterostructure constituted by metalated hydrogenated graphdiyne networks (hGDY) \cite{sun2016dehalogenative,yang2020metalated} and armchair graphene nanoribbons (aGNRs), as a prototypical model of hybrid graphene-GDY 2D heterostructure. By means of low-temperature scanning tunneling microscopy (LT-STM), we directly demonstrate the formation of covalent bonds at the hGDY–aGNR interface. Moreover, we show the possibility to control, through atomic hydrogen dosage, the surface density of Br atoms, which play a fundamental role in affecting the efficiency of the hGDY–aGNR covalent bonding. Density functional theory calculations elucidate the bonding mechanism at the hGDY–aGNR interface, providing a deep insight into the electronic properties of the heterostructure. Preliminary electronic transport calculations show the huge potential of hGDY-aGNR heterojunctions as new-generation electronic current switches.

\section{Results and discussion}\label{sec2}
\subsection{Structural identification of covalent hGDY-aGNR bonds}

\begin{figure}[h]
\centering
\includegraphics[width=1\textwidth]{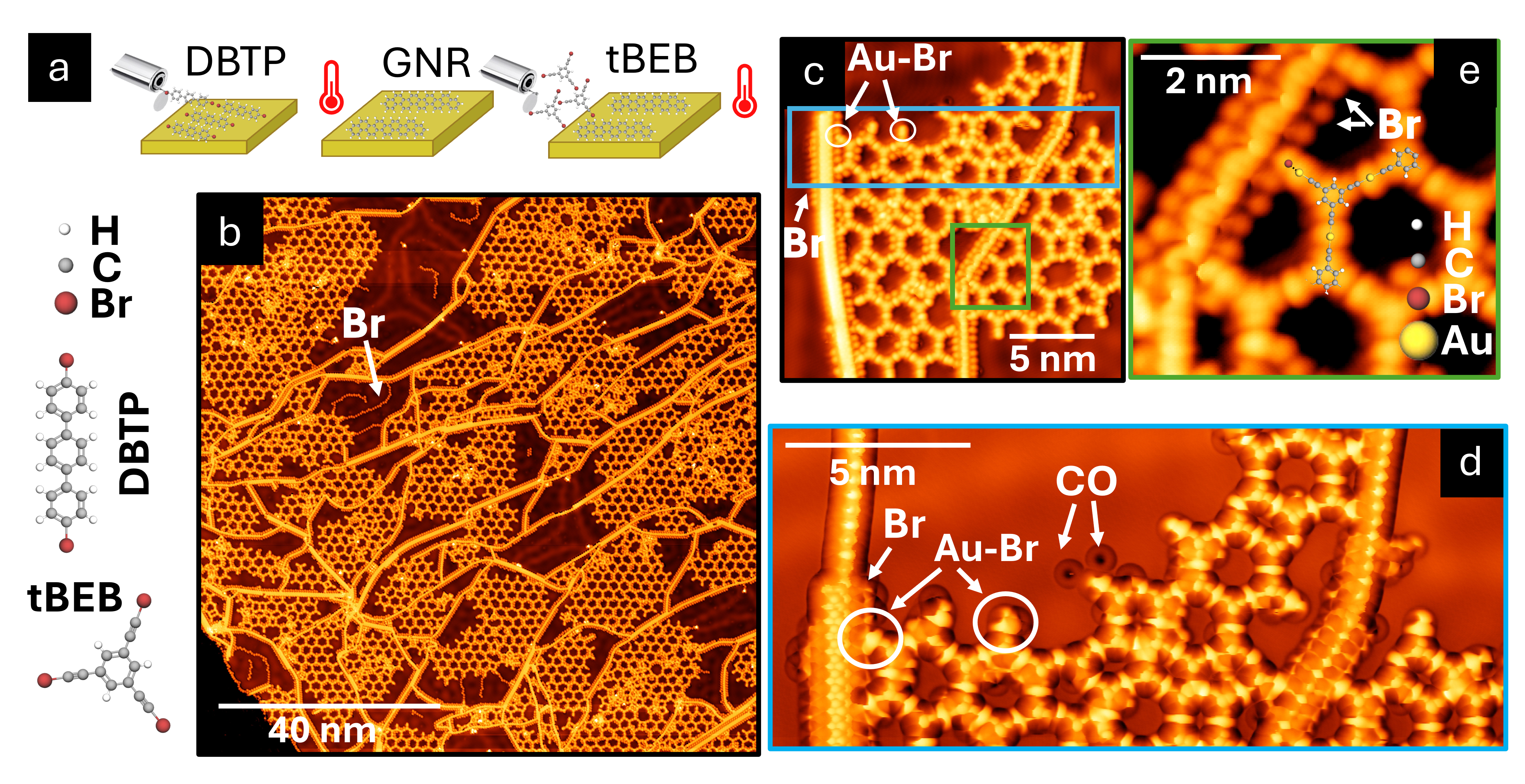}
\caption{(a) Schematic representation of the OSS of the hGDY–aGNR heterostructure on Au(111). On the left, the ball-and-stick atomic model of 4,4''-dibromo-p-terphenyl (DBTP) and 1,3,5-tri(bromoethynyl)benzene (tBEB) molecular precursors. Large-scale (b) and atomically-resolved CO-functionalized tip (c-e) LT-STM images of the as-deposited system on Au(111) before the hGDY–aGNR bonding. (d) Bond-resolved close-up of the squared blue region in (c). White circles in (c) and (d) mark Au-Br complex facing and not facing the nanoribbon. (e) Close-up of the squared green region in (c), showing an atomically-resolved LT-STM image of the Au-Br complex facing a 3-aGNR nanoribbon. The corresponding ball-and-stick atomic model is superimposed on the image.  STM setpoint: (b) -50mV/10pA, (c) -5mV/20pA, (d) -5mV/10pA, (e) -2mV/50pA.
}\label{OSS process & pre link}
\end{figure}

Lateral heterostructures between aGNRs and metalated hGDY are obtained via on-surface synthesis on Au(111) under UHV conditions, following a multi-step approach schematized in Figure \ref{OSS process & pre link}a (see Methods for growth details). 

Low coverage aGNRs samples (0.2-0.3 ML) are realized as scaffolds to host the growth of metalated hGDY in-between the nanoribbons: once the GNRs are obtained \cite{basagni2015molecules,merino2017width, de2020templating}, tBEB molecules are deposited on the surface decorated with the GNRs (see Figures S1-S2 and Methods). Upon annealing at 400 K to promote the ordering of the metalated hGDY, Br atoms coming from the dehalogenation of the hGDY precursor are found mainly along the edges of the aGNRs, being stabilized by the interaction with the H
atoms saturating the nanoribbons (see Figure \ref{OSS process & pre link}c-e and Methods). According to our interpretation, following this annealing step, Au adatoms terminating the metalated hGDY tend to form complexes with Br atoms decorating the nanoribbons (see Figure \ref{OSS process & pre link}c,e), as confirmed by ab initio calculations (see S3), preventing the formation of the hGDY–aGNR heterojunction. 

\begin{figure}[h]
\centering
\includegraphics[width=1\textwidth]{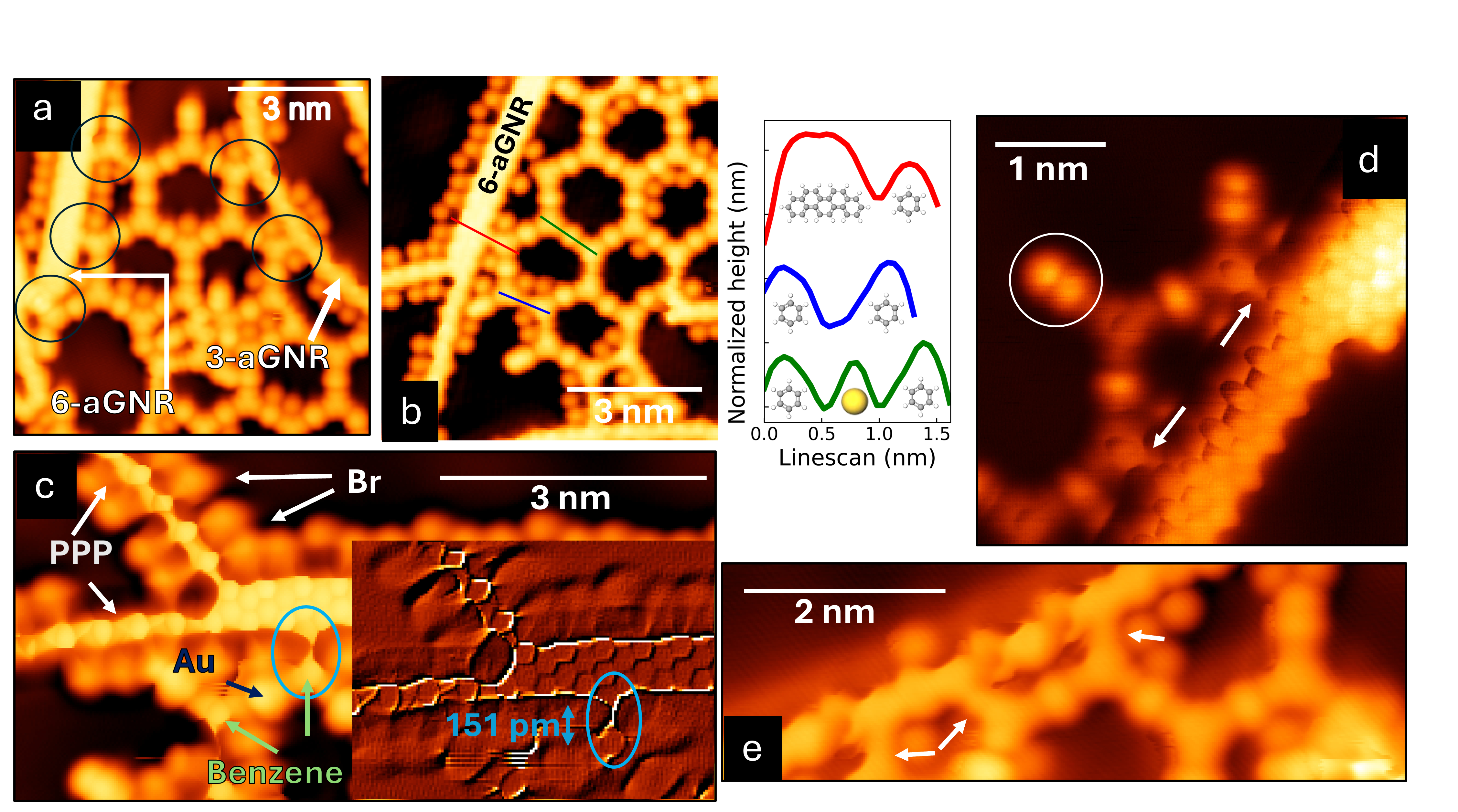}
\caption{(a)-(b) High resolution LT-STM images of two regions of the hGDY–aGNR heterostructure on Au(111) upon annealing at 530 K. The black circles in (a) exemplify the generated bonds between the hGDY and a 3- and a 6-aGNR. The line profiles taken along the blue, green and red lines on (b) are shown on the right, together with the ball-and-stick atomic model of the benzene ring and the gold adatoms. (c) Close-up, with CO-functionalized tip, of (b), rotated by 65º. The cyan circle encloses the hGDY–aGNR covalent bond. The inset shows the same LT-STM image as (c) processed with convolution filtering, with the cyan circle enclosing the hGDY–aGNR covalent bond. (d)-(e) Constant-height and constant-current LT-STM images of two junction points between hGDY and a 6- and 3-aGNR, respectively. The white circle in (d) exemplifies an Au-Br complex not facing the nanoribbon. White arrows in (d)-(e) indicate the position of the covalent hGDY–aGNR bonds. STM setpoint: (a) -10mV/80pA, (b) -10mV/80pA, (c) -3mV/80pA, (d) constant height at 0V, (e) 1mV/200pA.
}\label{annealing for link}
\end{figure}

Increasing the annealing temperature to 530 K enables the formation of covalent bonds between the metalated hGDY and aGNRs, as it can be seen in Figure \ref{annealing for link}a, where the hGDY is bonded to both 3- and 6-aGNRs. Statistical analysis performed over several LT-STM images of the heterostructures reveals that 83$\%$ of the covalent bonds form an angle of 90° between the acetylene units of the hGDY and the nanoribbon axis (Figure\ref{annealing for link}b, d and model ``A” in Figure S4), while only 17$\%$ form an angle of 60° (or, equivalently, 120°) (Figure \ref{annealing for link}e and model ``B” in Figure S4).

Figure \ref{annealing for link}b shows a high-resolution image of the hGDY–aGNR heterostructure upon formation of hGDY–aGNR bonds. Height profiles show a modulation along the green line, typical of the hGDY metalated phase, that disappears along the blue and red lines, indicating the removal of the Au adatom and C-C homocoupling, demonstrating that hGDY–aGNR connection is not mediated by a gold adatom, in agreement with previous observation on covalent GDY-like networks \cite{sun2016dehalogenative,rabia2019scanning,rabia2020structural,d2025unraveling,cartoceti2025surface}. 
Importantly, the bond-resolved imaging shown in the inset of Figure \ref{annealing for link}c yields a measurable length between the hGDY and the aGNR of approximately 151 pm, corresponding to a single C-C bond \cite{allen1987tables}. The same evidence was observed in several LT-STM images of other hGDY–aGNR bonds in different regions of the system (Figure S5).
Notably, the formation of hGDY–aGNR bonds occurs concurrently with the disappearance of the Au-Br complexes at the frontier between the two, as deduced by comparing the system before (Figure \ref{OSS process & pre link} and Figure S6a) and after (Figure \ref{annealing for link} and Figure S6b) the annealing at 530 K.

\begin{figure}[h]
\centering
\includegraphics[width=1\textwidth]{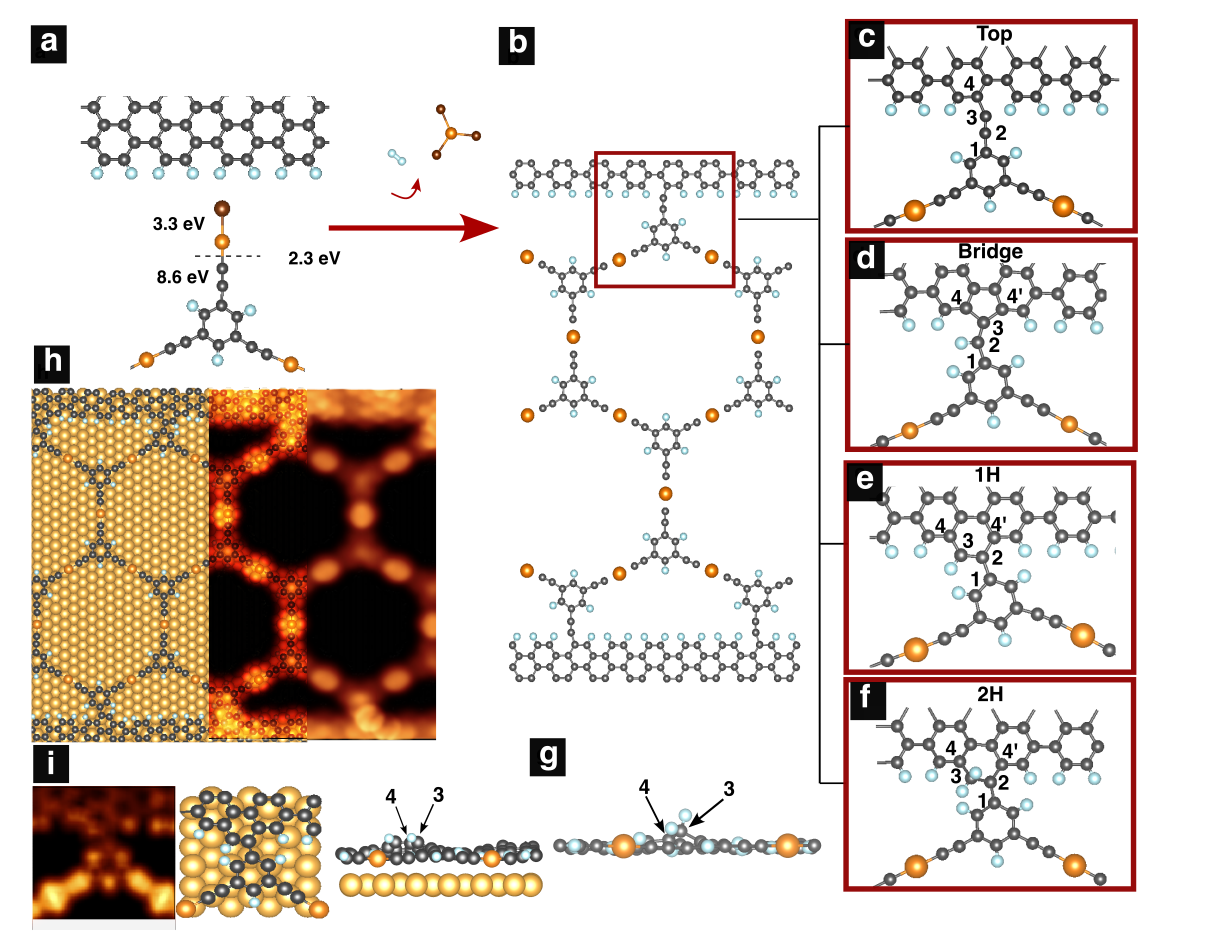}
\caption{a, b) hGDY-aGNR bonding mechanism: rupture of C-AuBr at 530 K in AuBr-terminated free edges of hGDY and consequent desorption of AuBr complexes; formation of a covalent bond with the nanoribbon and release of molecular hydrogen. Bond dissociation energies are reported in (a), together with the most probable breaking point (dashed). (c-f) Simulated model structures (close-up of the interface, as in red square of panel b) for the freestanding lateral heterostructure. Nomenclature of the carbon atoms at the interface adopted for the evaluation of the bond lengths reported in Table 1. g) Vertical distortion of the heterojunction in the most stable freestanding configuration (``2H''). h) Structural model for the gold-supported heterostructure in the ``1H'' configuration, together with the constant height (2 \AA) STM simulation for filled states (integration range [E$_F-0.5$ eV, E$_F$]. i) Enlargement of the bonding site: STM simulation at constant current and structural model (top and lateral view) for ``1H'' configuration on Au.}\label{models-freestanding}
\end{figure}

Based on these observations, we can deduce that the formation of a covalent bond between the two subsystems occurs through the thermally-activated breaking of the C-Au bond terminating the diacetylenic group of the metalated hGDY, and the consequent removal of the Au-Br complex at the hGDY–aGNR interface. Indeed, according to our calculations, the Au-Br dissociation energy (3.40 eV) is larger than the one obtained for the C-Au bond in the ethynyl moiety of the hGDY (2.26 eV) and smaller than the one of the $\mathrm{C\!\equiv\!C}$ ($\sim$ 8 eV) \cite{kass2024pi}. This suggests the C-Au bond as the most likely breaking point within the hGDY during the annealing at 530 K, as exemplified in Figure \ref{models-freestanding}a.

To clarify the nature of the bonding at the hGDY-aGNR interface, we performed ab initio calculations of the freestanding hGDY–aGNR heterostructure, considering different possible
connections
between 6-aGNR (i.e., the most abundant found in our samples)
and hGDY, as reported in Figure \ref{models-freestanding}c-f. 
Based on the statistics of the hGDY orientation deduced from STM images, we simulate a configuration with hGDY connected perpendicularly to 6-aGNR. The unit cell of the calculation was set as detailed in Methods.
In the structure reported in Figure \ref{models-freestanding}c, identified as ``Top'', the terminal $sp$ carbon atom of the hGDY (C$_3$ in the figure) and a de-hydrogenated carbon atom of the aGNR (C$_4$ in the figure) are vertically connected. In the "Bridge" configuration (panel d), the covalent bond is established between the terminal $sp$ carbon atom of the hGDY (C$_3$) and two carbon atoms of the aGNR (C$_4$ and C$_4'$) that have lost the saturating H atoms.

In panel e,f, both carbon atoms of the hGDY terminal acetylenic unit (C$_2$ and C$_3$) reorient, forming bonds with atoms C$_4$ and C$_4'$ at the edge of the aGNR, resulting in a cyclic structure.
The carbon atom C$_3$ is saturated with one (configuration ``1H'', Figure \ref{models-freestanding}e) or two (configuration ``2H'', Figure \ref{models-freestanding}f) H atoms, ensuring the tetravalency of carbon.
The total energy difference with respect to the most stable case for the freestanding heterostructure, i.e. the ``2H'', is reported in Table \ref{Tab1}, together with the bond lengths for selected pairs of atoms (C$_i$--C$_j$, $i=1,4$), testifying different carbon hybridization states depending on the explored configuration.

\begin{table}[htbp]
\centering
\caption{\textbf{Relative energies and structural parameters for ``Top'', ``Bridge'', ``1H'' and ``2H'' configuration shown in Figure \ref{models-freestanding}, both freestanding and on Au(111). Energies are in eV, bond lengths are in Angstrom.}}
\label{tab:parameters}
\begin{tabular}{lcccc||cc}
\toprule
 & \textbf{Top} & \textbf{Bridge} & \textbf{1H} & \textbf{2H} & \textbf{1H on Au} & \textbf{2H on Au} \\
\midrule
$E - E_\mathrm{stable}$ & 4.52 & 2.39 & 0.77 & 0.00 & 0.000 & 0.014 \\
C$_1$--C$_2$ & 1.42 & 1.46 & 1.53 & 1.46 & 1.48 & 1.42 \\
C$_2$--C$_3$ & 1.23 & 1.38 & 1.39 & 1.51 & 1.38 & 1.50 \\
C$_3$--C$_4$ & 1.42 & 1.52 & 1.42 & 1.49 & 1.41 & 1.49 \\
C$_3$--C$_4'$ & -- & 1.51 & -- & -- & -- & -- \\
C$_2$--C$_4'$ & -- & -- & 1.48 & 1.43 & 1.46 & 1.45 \\
$\Delta$z$_{3-4}$ & -- & 0.05 & 0.09 & 0.51 & 0.08 & 0.51 \\
$\Delta$z$_{3-\mathrm{GNR}}$ & -- & 0.08 & 1.23 & 1.31 & 1.14 & 1.46 \\
\bottomrule
\end{tabular}
\label{Tab1}
\end{table}

In the ``Top'' model, the $sp$-hybridization of the C$_2$-C$_3$ diacetylenic unit terminating the hGDY is preserved. Differently, in the ``Bridge'', two single bonds are formed between the terminal C$_3$ atom of the hGDY and the C$_4$ and C$_4'$ atoms of the aGNR, resulting in a fulvene ring. This modifies the hybridization state of C$_2$ and C$_3$ atoms from $sp$ to $sp^2$.
In the ``1H'' configuration (panel e), the C$_3$-C$_4$ and C$_2$-C$_4'$ bond lengths are compatible with $sp^2$ hybridization; differently, in the ``2H'' (panel $f$) the hybridization of C$_3$ becomes $sp^3$, generating a local tetrahedral structure (Figure \ref{models-freestanding}g) with a large local atomic displacement, reported in terms of the height difference between atoms C$_3$ and C$_4$ ($\Delta$z$_{3,4}$) compared to the other three configurations. All the structures except the ``Top'' are anyway partially buckled, as can be deduced by the height difference  ($\Delta$z$_{3,GNR}$ in Table~\ref{Tab1}) between the C$_3$ atom and one carbon atom internal to the nanoribbon, where the structure is planar. The buckling is larger for ``1H'' and ``2H'' configurations.

Based on the total energy differences reported in Table \ref{Tab1}, we can infer that the formation process of the hGDY–aGNR heterojunction may take place through the initial formation of a ``Top'' configuration, i.e. the most likely due to the proximity of the ending carbon bond in the hGDY fragment and the aGNR, possibly followed by a distortion of the acetylene bond that drives the system in configuration ``1H'' or ``2H''. Indeed, the formation energy for the latter, calculated as detailed in Methods, is equal to -3.55 eV. 

To investigate the nature of the metal-supported heterojunction, we performed a calculation for the two most stable configurations in Figure \ref{models-freestanding}, i.e., ``1H'' and ``2H'' (panel e-f), on Au(111). Our model, reported in Figure \ref{models-freestanding}h, is built according to the experimentally observed orientation of the acetylene bonds of the hGDY along the Au(111) [11$\overline{2}$] direction \cite{rabia2020structural, sun2016dehalogenative}. The periodicity of the
heterostructure is rescaled to fit that of Au(111), resulting in a $\sim 2\%$ compression of the carbon network with respect to the freestanding one.
In the gold-supported system, we observe a mild flattening of the heterostructure due to the interaction with the metal substrate ($\Delta$z$_{3,GNR}$ in Table \ref{Tab1}, right). This leads to a stabilization of the ``1H'' configuration with respect to ``2H'', according to the formation of an aromatic ring at the bonding site (Figure \ref{models-freestanding}i). Although the energy difference between the two models (i.e., 14 meV) is too small to exclude the formation of the ``2H'' configuration, the agreement between the calculated C$_1$-C$_2$ distance and the one extracted from the experiments, i.e., 1.51 \AA~(see Figure \ref{annealing for link}c), suggests the ``1H'' configuration as the most likely, leading to the formation of an additional aromatic ring at the edge of the aGNR. Except for the interface region, both hGDY and aGNR remain almost planar after structural relaxation. The simulated STM image obtained in the energy interval [E$_\mathrm{F}$-0.5,E$_\mathrm{F}$] is in agreement with the main features observed experimentally, i.e., the height modulation between brighter and fainter spots, corresponding to Au adatoms and phenyl rings alternation, and the covalent interface bond (Figure \ref{models-freestanding}h-i).

\subsection{Role of Br atoms in the formation efficiency of hGDY–aGNR heterojunction}

hGDY-aGNR heterostructures obtained through the described method demonstrate an average efficiency of 47\% in the formation of hGDY–aGNR covalent bonds, with an average Br density of 0.53 nm$^{-2}$.
To disclose the role of Br atoms, we tested different experimental conditions, proving the possibility to control their surface density and improve the heterojunction formation efficiency. We first increased the average number of Br atoms per unit area to 0.84 nm$^{-2}$ by increasing the tBEB molecules deposition time from 20 to 30 minutes. This resulted in a decrease to 31\% of the hGDY–aGNR bonding efficiency. Then, we decreased the average number of Br atoms per unit area by dosing atomic hydrogen \cite{zuzak2020surface} (see Methods). By optimizing hydrogen gas pressure and dosage time, we reduced the average number of Br atoms on the surface to 0.19 nm$^{-2}$, resulting in an increased efficiency in the formation of the hGDY–aGNR covalent bonds up to 71\% (see Figures S7-S8), although complete Br removal was not achieved and some degradation and disordering effects possibly associated with the high reactivity of the diacetylenic bonds toward hydrogen were observed (see Figure S9).

\subsection{Electronic properties at the hGDY-aGNR interface}

\begin{figure}[h!]
\centering
\includegraphics[width=1.\textwidth]{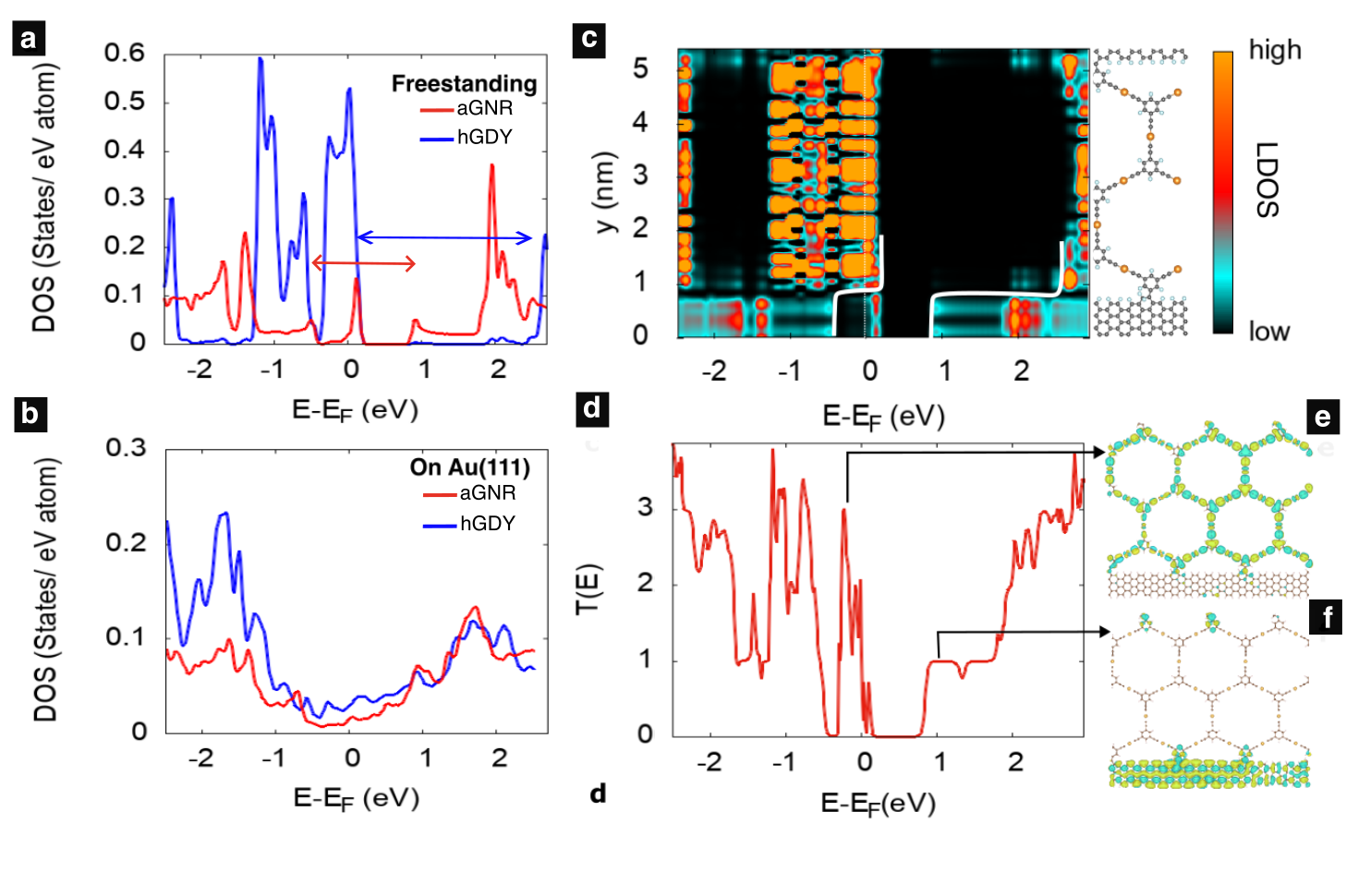}
\caption{PDOS on carbon atoms of 6-GNR (red) and hGDY (blue) in the freestanding ``2H'' (a) and gold-supported ``1H'' (b) heterojunction. c) Average total LDOS projected on all the atoms with the same y coordinates as a function of the lateral position along the freestanding ``2H'' heterojunction. The band alignment at the junction is evidenced by white lines that mark the edges of the gap in the two subsystems. d) Zero-bias transmission function of the freestanding ``2H'' heterostructure. PDOS and transmission of the freestanding ``1H'' heterojunction are reported in Figure S13. e) First conductive transmission eigenchannels at E=-0.2 eV and f) at E=1 eV. 
}\label{el-prop}
\end{figure}

Considering the relevance of 2D all-carbon lateral heterostructures in view of their potential application in organic electronics, we compared the electronic properties of the most stable freestanding hGDY-aGNR heterostructure (Figure \ref{models-freestanding}f) with its gold-supported counterpart (Figure \ref{models-freestanding}h). The calculated projected density of states (PDOS) for the freestanding and Au(111)-supported systems, resolved on the contributions of aGNR (red) and hGDY (blue) carbon atoms, are shown in Figure \ref{el-prop}a,b. Upon the formation of the heterostructure, the metalated hGDY retains its metallic character. In the energy gap of 2.3 eV just above the Fermi level
(blue arrow in Figure \ref{el-prop}a),
some features appear, due to states localized at the interface with aGNR, that decay in the core of the hGDY (Figure \ref{el-prop}c and blue line in Figure S10a).  In the aGNR, the energy gap of 1.2 eV is preserved (red arrow in Figure \ref{el-prop}a), and the positioning of the Fermi level with respect to the gap edges suggests a mild $p$-type doping of the aGNR, in analogy with the behaviour of aGNR on gold \cite{merino2017width}. The feature at the Fermi level is due to a state localized at the edge of the ribbon, in proximity of the bonding with the hGDY (Figure S10b). From the decay length of the aGNR and hGDY interface states, as inferred from Figure \ref{el-prop}c, the extension of the electronic junction is estimated to be approximately 1 nm.

The electronic abruptness of the junction allows the two subsystems to preserve their own features in the DOS, as it emerges from a comparison with the PDOS of the pristine, non-connected hGDY and aGNR (Figure S11a).

In the gold-supported heterostructure (Figure \ref{models-freestanding}g), the aGNR and hGDY PDOS (Figure \ref{el-prop}b) display the fingerprint of the hybridization with the substrate states and the Fermi level shift towards higher energies, as already observed for the single hGDY \cite{rabia2020structural} and aGNR on Au(111) \cite{merino2017width}.
Their similarity with the PDOS of isolated subsystems on Au(111) (Figure S11b) confirms that the main interaction occurs with the gold surface. Nevertheless, it is worth noticing that there is a net charge transfer from hGDY to the aGNR ($\sim 0.5$ electrons per aGNR unit), as deduced from a Mulliken charge analysis, leading to a reduction of the $p$-doping of the aGNR on Au(111).

Although the calculated electronic properties show a relevant effect of the metallic substrate on the properties of the supported heterostructure, based on the results obtained for the freestanding system, we can infer that its transfer or direct growth on semiconducting/insulating supports would result in a system in which the hGDY and aGNR are electronically decoupled. 

To evaluate the transport properties of the heterojunction, we analyzed the zero-bias transmission (T(E), see Methods) of the freestanding heterostructure, mimicking its behaviour on a non-metal substrate. The calculated T(E), reported in Figure \ref{el-prop}d, does not display the step-like behaviour typical of systems with perfectly conductive transmission channels (T(E)=$\sum_i t_i(E)$ with $t_i$=1). This is due to the intrinsic disorder induced by local distortions at the interface, where the hGDY connect to aGNR via an H-saturated carbon atom. However, depending on the energy, one or more transmission channels are activated, and they can be associated with states of the hGDY, aGNR, or both (see Figure \ref{el-prop}a and c). 
To analyze the contribution of the two subsystems to the transmission, we calculate the transmission eigenchannels from left to right, i.e. the eigenvectors of the transmission matrix, at -0.2 eV and 1 eV (arrows in Figure \ref{el-prop}d), and we report the spatial dependency of the first eigenchannel in Figure \ref{el-prop}e and f.

At  E=-0.2 eV, there are four activated eigenchannels with transmission probability $\sim 1$. They show a spatial distribution in the hGDY (see Figure \ref{el-prop}e for the first eigenchannel and Figure S12 for the other three), demonstrating that by applying a small bias, the current would flow only in the hGDY.
Differently, at E=1 eV, only one channel contributes to the T(E), and involves only states of the aGNR \ref{el-prop}f.
Based on these results, a hypothetical device could exploit a suitable combination of source–drain and gate voltages to tune the Fermi level and selectively drive the current through specific regions of the structure, in analogy with predictions for other organic junctions \cite{H-dimers}.
Although this analysis deserves further investigations and experimental proof of concept, it is an example of the huge potential of covalently bonded hGDY–aGNR lateral heterostructures.

\newpage
\section{Conclusions}
To conclude, we unveiled the synthesis of a covalently bonded hybrid $sp$-$sp^2$ lateral heterostructure between graphene nanoribbons and extended metalated graphdiynes on Au(111). Combining LT-STM imaging and DFT calculations, we provided a deep insight into the mechanism of formation of the hGDY–aGNR covalent bond, inferring its sp$^2$ character. Notably, we found a preferential bonding direction of the hGDY perpendicular to the GNRs in registry with the underlying substrate. We demonstrated the major role played by chemisorbed Br atoms in affecting the formation efficiency of a covalent hGDY–aGNR heterojunction, that reaches a value of 71\% through controlled atomic hydrogen dosage. 
Simulated electronic properties of the freestanding heterostructure show an electronically abrupt heterojunction with the two subsystems preserving their own DOS features upon formation of the heterojunction, while they evidence the effect of the hybridization with the substrate when metal-supported.
For the freestanding heterostructure, we propose a possible application for current spatial separation in organic junction-based devices.

This work demonstrates the potential of hGDY in synergically complementing graphene nanoribbons as a prototype of graphene-GDY heterostructures for new-generation nanoelectronics, with huge downscaling potential to atomic-scale circuitry.

\newpage
\section{Methods}\label{sec11}

\subsection*{Experiment}
All the experiments were carried out under ultra-high vacuum conditions, with a base pressure below $10^{-10}$~mbar. Single-crystal Au(111) (MaTeck GmbH) was employed for the on-surface synthesis process, following several cleaning cycles of Ar$^+$ sputtering and annealing at 770~K. Molecular precursors, i.e. 1,3,5-tri(bromoethynyl)benzene (tBEB), and 4,4''-dibromo-p-terphenyl (DBTP), were loaded in powder form in two separate Knudsen cells and were deposited on Au(111) by thermal evaporation, after calibration of the molecular flux with a quartz microbalance.
The lateral heterostructure was obtained through the sequential deposition of DBTP and tBEB molecules on Au(111) substrate, held at room temperature in front of the evaporator. To firstly obtain aGNRs, DBTP molecules are sublimated on Au(111) for 10 minutes, with a rate of about 0.025 ML/minute, keeping the crucible temperature set to 353~K, reaching a chamber pressure of 6~$\times~10^{-10}$~mbar. Upon evaporation at RT, DBTP molecules form an ordered self-assembly of unreacted monomers stabilized by lateral hydrogen bonds between Br and H terminals \cite{basagni2015molecules} (see Figure S1a). It is known that the debromination of DBTP molecules occurs in the 360-400 K range, and Br desorption from Au(111) surface at about 650 K \cite{basagni2015molecules}. Upon annealing at 523 K for 30 minutes, poly(p-phenylene) (PPP) chains are formed \cite{basagni2015molecules} (see Figure S1b). Further annealing of the sample at 773 K for 30 minutes promotes the cyclodehydrogenation of PPP chains, and their coupling into aGNRs (see Figure S1c) with polydisperse widths, from 3- to 6- and 9-aGNR \cite{merino2017width}, and interconnecting branches (see Figure S2). tBEB molecules are evaporated for 20 or 30 minutes, depending on desired coverage, over the obtained aGNR/Au(111) sample, keeping the crucible at 303 K, reaching a chamber pressure of 5~$\times~10^{-10}$~mbar. Upon surface-catalyzed halogen cleavage, tBEB molecules undergo coupling reaction through the incorporation of gold adatoms from the substrate, generating metalated hGDY in-between the nanoribbons \cite{d2025unraveling, sun2016dehalogenative, yang2020metalated, shu2020atomic, rabia2020structural}. To improve the order of the hGDY, a mild annealing at about 400 K is performed.
Br atoms are found both inside the pores of the hGDY hexagonal network and outside, decorating the nanoribbons, with a much higher statistical preference for the latter configuration. In fact, in Figure \ref{OSS process & pre link}b,c, it is possible to observe the almost complete decoration of both sides of the nanoribbons by Br atoms, while the average number of Br atoms inside a single pore of the hexagonal network is 2, much lower than the one found for the same system in the absence of the aGNRs, namely 3.5 on Au(111) and 5.1 on Ag(111) \cite{d2025unraveling}. Further indication of Br tendency to interact with hydrogen can be found in the nucleation of the 2D metalated hGDY, which occurs just in the proximity of the aGNRs. This can be appreciated in Figure S14, showing the STM image of a low tBEB coverage aGNR/Au(111) sample.
To promote the bonding between the metalated framework and the graphene nanoribbons, the hGDY-aGNR/Au(111) sample is annealed for 15 min at 530 K. 

Low-temperature scanning tunneling microscopy (LT-STM) investigations were conducted at the Laboratorio de Microscopías Avanzadas of the Universidad de Zaragoza, with a Scienta Omicron microscope, cooled down to 4.8 K, using a W tip. Measurements of bond lengths have been performed on LT-STM images featuring a CO functionalized tip and processed with directional convolution filtering on Gwyddion software \cite{nevcas2012gwyddion}. In particular, two custom 3×3 directional kernels have been applied to obtain features sharpening and atomic corrugation enhancement: the first kernel ([[-1, 0, 0], [0, 2, 0], [0, 0, -1]]) and the second kernel ([[0, 0, -1], [0, 2, 0], [-1, 0, 0]]). The final image was obtained by averaging the results of both convolutions.

Hydrogenation was carried out using a home-built hydrogen cracker with a hot filament, according to the procedure already reported in \cite{zuzak2020surface}. A leak valve connected to the hydrogen gas line allowed the desired dosage of atomic hydrogen on the sample surface. During hydrogenation, the sample was kept at room temperature, perpendicular to the cracker. The filament current of the cracker was set to 4 A, and the chamber pressure during hydrogen dosage was 2~$\times~10^{-8}$~mbar. Hydrogenation was carried out upon the formation of hGDY-GNR bonds, i.e. upon annealing at 530 K; subsequently to the H dosage, another annealing at the same temperature (i.e., 530 K) was performed.

Statistics of the surface density of Br atoms was done over several STM images taken in different regions of the hGDY-GNR/Au(111) sample upon the formation of hGDY-GNR bonds. In particular, for the "standard" samples, with the tBEB deposition time set to 20 minutes and no H dosage, the statistics were done upon sample annealing at 530 K, i.e. upon formation of hGDY-GNR bonds. On the samples obtained with the same tBEB deposition time, thus the same starting amount of Br atoms, but with H dosage, the statistics was done upon H dosage and consequent annealing at 530 K. Finally, for the samples with tBEB deposition time set to 30 minutes, i.e. with a higher starting amount of Br atoms, the statistics was done upon sample annealing at 530 K. To summarize, the statistics was done on every sample after the same annealing step, i.e. the one leading to the hGDY-GNR bonds formation (530 K).

\subsection*{Theory}
Theoretical calculations have been performed in density functional theory by exploiting a pseudopotential description of electron-ion interaction and atomic orbitals basis set as implemented in the SIESTA code. We adopted a mesh cutoff of 400 Ry, the GGA-PBE exchange and correlation functional and we used Grimme dispersion forces to include the van der Waals correction for the interaction between the carbon heterostructure and the system.
All the structures have been prepared following the criterion of minimum stress.
To limit the size of the system and exploit periodic boundary conditions, we consider the unit cell reported in Figure \ref{models-freestanding}b as an example of the heterostructure. For all the simulated configurations, structural relaxation and cell optimization were performed. In the starting configuration, the hGDY lattice constant is adapted to the aGNR one, and the optimization leads to a final minimum-energy configuration characterized by a residual stress between the two subsystems smaller than 1\% with respect to the freestanding ones. Indeed, when periodic boundary conditions are imposed on two subsystems with a lattice mismatch, it is not possible to obtain a stress-free commensurate structure in both components. Nevertheless, the almost negligible structural stress resulting from our calculation makes this description reliable.

Considering the formation mechanism reported in Figure \ref{models-freestanding}a, the binding energy per covalent bond formed at the hGDY–aGNR interface is estimated as
$$E_B=E_{GNR}+E_{GDY-AuBr}+E_{Br_2}-E_{GNR+GDY}+E_{AuBr_3}+nE_{H_2}$$
with $E_{GNR}$ and $E_{GDY-AuBr}$ the energy of the isolated aGNR and AuBr-terminated hGDY, $E_{Br_2}$, E$_{AuBr_3}$ and $nE_{H_2}$ the energies of the molecules desorbed in the process, and $E_{GNR+GDY}$ the total energy of the heterostructure.

For the Au(111)-supported heterostructure, the periodicity of the hGDY and aGNR has been adapted to that of the gold surface, leading to a compression of 3.3\% and 3.5\% for the hGDY and aGNR, respectively. 
STM images are generated in Tersoff-Hamann approach at costant distance of 3 a.u. from the surface.
Electronic transport was computed using the Non-Equilibrium Green’s Function (NEGF) formalism as implemented in \textsc{TranSIESTA}, the DFT-based transport module of \textsc{SIESTA}~\cite{Brandbyge2002}. The system was divided into two semi-infinite electrodes and a central scattering region. The effect of each electrode was included through the \textit{self-energies} $\Sigma_{L,R}^{r}(E)$, which account for the open-boundary coupling between the finite device region and the infinite leads, thereby embedding the electrodes into the calculation. The retarded Green’s function was evaluated as
\[
G^{r}(E)=\left[E S - H - \Sigma_L^{r}(E) - \Sigma_R^{r}(E)\right]^{-1},
\]
and the transmission function was obtained as
\[
T(E)=\mathrm{Tr}\!\left[\Gamma_L(E)\, G^{r}(E)\, \Gamma_R(E)\, G^{a}(E)\right],
\qquad
\Gamma_{L,R}(E)=i\left[\Sigma_{L,R}^{r}(E)-\Sigma_{L,R}^{a}(E)\right].
\]
Transport eigenchannels were computed by diagonalizing
\[
\mathbf{t}^\dagger(E)\,\mathbf{t}(E),
\]
where $\mathbf{t}(E)$ is the transmission matrix derived from $G^{r}(E)$ and the electrode–device couplings.
We adopt a k-mesh of $9\times4\times1$ for the electrode calculation, $3\times4\times1$
for the transiesta calculation and $1\times10\times1$ for the calculation of the current in TBtrans (current in the horizontal direction).
The transmission probability is expressed in units of quantum conductance (G$_0$ = 2e$^2$ /h = 77.48 mS).

\section{Data availability}
The data supporting the findings of this study are available in the Article and its Supplementary Information. Additional data are available from the corresponding authors on request.


\section{Acknowledgments}
A.C., A.L.B. and C.S.C acknowledge funding by: Funder: project funded under the National Recovery and Resilience Plan (NRRP), Mission 4 Component 2 Investment 1.3 - Call for tender No. 341 of 15.03.2022 of Ministero dell'Universit\'a e della Ricerca (MUR); funded by the European Union NextGenerationEU - Award Number: project code PE0000021, Concession Decree No. 1561 of 11.10.2022 adopted by Ministero dell’Universit\'a e della Ricerca (MUR), CUP D43C22003090001, Project title ``Network 4 Energy Sustainable Transition NEST' '. F.S. acknowledges the financial support from the University of Padova through the grant  P-DiSC\#02BIRD2024-UNIPD  VS-PpyTM. J.L.C. acknowledges financial support from the Spanish Ministry of Science and Innovation (Grant PID2022\_138750NB-C21 and ``Severo Ochoa' ' Programme for Centres of Excellence in R\&D CEX2023-001286-S funded by MCIN/AEI/ 10.13039/501100011033 and by ERDF ``A way of making Europe' ' ), from the regional Governments of Aragon (E12\_23R) and from the European Commission (project ULTIMATE-I, grant ID 101007825). J.L.C. and A.E.C. further acknowledge the use of Servicio General de Apoyo a la Investigaci\'on-SAI and the Laboratorio de Microscopías Avanzadas of the Universidad de Zaragoza. S.A. acknowledges the CINECA award under the ISCRA initiative , for the availability of high-performance computing resources and support, and the financial support from INFN (Progetto iniziativa specifica ``TIME2QUEST''). S.A. also acknowledges Federico Civaia for performing preliminary calculations on systems similar to those presented in the manuscript. A.O.B. acknowledges support from the Ministero dell’Universit\'a e della Ricerca (MUR) and the University of Pavia through the program ``Dipartimenti di Eccellenza'' 2023--2027.

\section{Author contributions}
A.C. conceived and conducted the STM experiments, analyzed the corresponding data, and wrote the initial draft. S.A. conceived the theoretical modelling, performed the DFT calculations, analyzed the corresponding data and wrote the initial draft.
A.C., A.E.C and J.L.C. conducted LT-STM experiments. A.C. and P.D. performed preliminary RT-STM measurements.
M. A. and E. B. contributed to DFT calculations. A.O.B. synthesized the tBEB molecular precursor. J.L.C., A.L.B., C.S.C. contributed to the supervision and discussion of results. A.C. and S.A. contributed equally to this work. All authors contributed to the revision and final discussion of the manuscript.

\section{Competing interests}
The authors declare no competing interests.

\end{document}